# Millennium Pathways for Tractography:
# 40 grand challenges to shape the future of tractography

Maxime Descoteaux[1*], Kurt G. Schilling[2,3*], Dogu Baran Aydogan[4], Christian Beaulieu[5], Elena Borra[6], Maxime Chamberland[7], Alessandro Daducci[8], Alberto De Luca[9], Flavio Dell'Acqua[10], Jessica Dubois[11,12], Tim B. Dyrby[13,14], Shawna Farquharson[15], Stephanie Forkel[16,17,18], Martijn Froeling[19], Alessandra Griffa[20,21], Mareike Grotheer[22,23], Pamela Guevara[24], Suzanne N. Haber[25,26], Vinod Kumar Jangir[27], Alexander Leemans[28], Joël Lefebvre[29], Ching-Po Lin[30], Graham Little[31], Chun-Yi Zac Lo[32], Chiara Maffei[33], Helen S. Mayberg[34], Jennifer A. McNab[35], Pratik Mukherjee[36], Lauren J. O'Donnell[37,38], Martin Parent[39], Carlo Pierpaoli[40], Francois Rheault[1], Kathleen S. Rockland[41], Alard Roebroeck[42], Ariel Rokem[43], R. Jarrett Rushmore[44], Silvio Sarubbo[45], Simona Schiavi[46,47], Stamatios N Sotiropoulos[48,49], Diego Szczupak[50], Michel Thiebaut de Schotten[51,52], J-Donald Tournier[53], Francesco Vergani[54], Joseph Yuan-Mou Yang[55,56,57], Fan Zhang[58], Derek Jones[59#], Laurent Petit[60,61#]

[1]Sherbrooke Connectivity Imaging Lab (SCIL), Computer Science department, Université de Sherbrooke, [2]Department of Radiology, Vanderbilt University Medical Center, Nashville, TN, USA, [3]Vanderbilt University Institute of Imaging Science, Vanderbilt University Medical Center, Nashville, TN, USA, [4]University of Eastern Finland, [5]Department of Radiology and Diagnostic Imaging & Biomedical Engineering, University of Alberta, Edmonton, Alberta, Canada, [6]Unit of Neuroscience, Department of Medicine and Surgery, University of Parma, Italy, [7]Department of Mathematics and Computer Science, Eindhoven University of Technology, Eindhoven, The Netherlands, [8]Computer Science Department, University of Verona, Strada le Grazie 15, Verona, 37134, Italy, [9]Image Sciences Institute, university medical center Utrecht, Utrecht, The Netherlands, [10]Department of Neuroimaging, Institute of Psychiatry, Psychology and Neuroscience, King's College London, London, UK, [11]Université Paris Cité, Inserm, NeuroDiderot, F-75019 Paris, France, [12]Université Paris-Saclay, CEA, NeuroSpin, UNIACT, F-91191, Gif-sur-Yvette, France, [13]Danish Research Centre for Magnetic Resonance, Department of Radiology and Nuclear Medicine, Copenhagen University Hospital Amager and Hvidovre, Hvidovre, Denmark, [14]Department of Applied Mathematics and Computer Science, Technical University of Denmark, Kongens Lyngby, Denmark, [15]The University of Sydney, [16]Donders Institute for Brain Cognition Behaviour, Radboud University, Nijmegen, The Netherlands, [17]Max Planck institute for psycholinguistics, Nijmegen, The Netherlands, [18]Brain Connectivity and Behaviour Laboratory, Sorbonne Universities, France, [19]Center for Image Sciences, Precision Imaging Group, Division Imaging & Oncology, University Medical Centre Utrecht, Utrecht, The Netherlands, [20]Leenaards Memory Centre, Department of Clinical Neurosciences, Lausanne University Hospital and University of Lausanne, Chemin de Mont-Paisible 16, 1011, Lausanne, Switzerland, [21]Neuro-X Institute, École Polytechnique Fédérale De Lausanne, Chemin des Mines 9, 1202, Geneva, Switzerland, [22]Department of Psychology, Philipps-Universität Marburg, Schulstraße 12, 35037 Marburg, Germany, [23]Center for Mind, Brain and Behavior - CMBB, Universities of Marburg, Gießen, and Darmstadt, Germany, [24]Faculty of Engineering, Universidad de Concepción, Concepción, Chile, [25]University of Rochester School of Medicine, Rochester, NY, [26]McLean Hospital, Harvard Medical School, Belmont, MA, [27]Max Planck Institute for Biological cybernetics, Tuebingen, Germany, [28]PROVIDI Lab, Image Sciences

Institute, UMC Utrecht, Utrecht, The Netherlands, [29]Laboratoire d'imagerie numérique, neurophotonique et microscopie (LINUM) Université du Québec à Montréal (UQAM), [30]Institute of Neuroscience, National Yang Ming Chiao Tung University, Taipei 112, Taiwan, [31]Department of Computer Science, Université de Sherbrooke, Sherbrooke, Quebec, J1K 2R1, Canada, [32]Institute of Intelligent Bioelectrical Engineering, National Yang Ming Chiao Tung University, [33]Athinoula A. Martinos Center for Biomedical Imaging, Massachusetts General Hospital and Harvard Medical School, Boston, MA, USA Center for Neurotechnology and Neurorecovery, Department of Neurology, Massachusetts General Hospital, Boston, MA, USA, [34]Nash Family Center for Advanced Circuit Therapeutics Icahn School of Medicine at Mount Sinai 1000 10th Ave. New York NY 10019 USA, [35]Department of Radiology, Stanford University, Stanford, CA, USA, [36]University of California, San Francisco, [37]Brigham and Women's Hospital, [38]Harvard Medical School, [39]Université Laval CERVO Brain Research Centre, [40]Laboratory on Quantitative Medical Imaging, National Institute of Biomedical Imaging and Bioengineering, NIH, [41]Department of Anatomy&Neurobiology, Chobanian&Avedisian School of Medicine, Boston University, 72 East Concord St., Boston, MA 02118 USA, [42]Department of Cognitive Neuroscience, Faculty of Psychology & Neuroscience, Maastricht University, Maastricht, the Netherlands, [43]Department of Psychology and eScience Institute, University of Washington, Seattle, WA, USA, [44]Department of Anatomy and Neurobiology, Boston University School of Medicine, [45]Center for Medical Sciences (CISMed), University of Trento, Provincia Autonoma di Trento, Italy; Department of Neurosurgery, Azienda Provinciale per i Servizi Sanitari (APSS), Provincia Autonoma di Trento, Italy, [46]ASG Superconductors S.p.A., Genoa, Italy, [47]Department of Computer Science, University of Verona, Verona, Italy, [48]Sir Peter Mansfield Imaging Centre, Mental Health and Clinical Neurosciences, School of Medicine, University of Nottingham, UK, [49]NIHR Nottingham Biomedical Research Centre, Nottingham University Hospitals NHS Trust, Queen's Medical Centre, Nottingham, UK, [50]University of Pittsburgh Brain Institute, Department of Neurobiology, University of Pittsburgh, 3501 Fifth Avenue, Pittsburgh, PA 15261, USA., [51]Groupe d'Imagerie Neurofonctionnelle, Institut des Maladies Neurodégénératives-UMR 5293, CNRS, CEA, University of Bordeaux, Bordeaux, France, [52]Brain Connectivity and Behaviour Laboratory, Paris, France, [53]Department of Early Life Imaging, School of Biomedical Engineering and Imaging Sciences, King's College London, London, UK, [54]King's College Hospital, London, UK, [55]Department of Neurosurgery, Neuroscience Advanced Clinical Imaging Service (NACIS), The Royal Children's Hospital, 50 Flemington Road, Melbourne, 3052, Victoria, Australia., [56]Neuroscience Research, Murdoch Children's Research Institute, 50 Flemington Road, Melbourne, 3052, Victoria, Australia, [57]Department of Paediatrics, University of Melbourne, 50 Flemington Road, Melbourne, 3052, Victoria, Australia, [58]University of Electronic Science and Technology of China, [59]Cardiff University Brain Research Imaging Centre (CUBRIC), School of Psychology, Cardiff University., [60]Université Bordeaux, CNRS, CEA, IMN, GIN, UMR 5293, F-33000 Bordeaux, France IRP OpTeam, CNRS Biologie, France - Université de Sherbrooke, Canada, [61]IRP OpTeam, CNRS Biologie, France - Université de Sherbrooke, Canada, [*]co-first author, [#]co-last author

# Abstract

In the spirit of the historic *Millennium Prize Problems* that heralded a new era for mathematics, the newly formed International Society for Tractography (IST) has launched the *Millennium Pathways for Tractography*, a community-driven roadmap designed to shape the future of the field. Conceived during the inaugural *Tract-Anat Retreat*, this initiative reflects a collective vision for advancing tractography over the coming decade and beyond. The roadmap consists of 40 grand challenges, developed by international experts and organized into seven categories spanning three overarching themes: neuroanatomy, tractography methods, and clinical applications. By defining shared short-, medium-, and long-term goals, these pathways provide a structured framework to confront fundamental limitations, promote rigorous validation, and accelerate the translation of tractography into a robust tool for neuroscience and medicine. Ultimately, the *Millennium Pathways* aim to guide and inspire future research and collaboration, ensuring the continued scientific and clinical relevance of tractography well into the future.

# The Origins of the International Society for Tractography and the First Tract-Anat Retreat

The *Millennium Pathways for Tractography* emerged from a growing realization within the tractography community: while the field had matured into a globally distributed, interdisciplinary network, it lacked the necessary formal structure to sustain long-term collaboration, coordination, and progress. This insight became evident during the development of the *Handbook of Diffusion MR Tractography (Dell'Acqua et al. 2024)*, where contributions from over a hundred researchers underscored both the depth of expertise and the breadth of interest across anatomical, technical, and clinical domains.

Recognizing this need for a cohesive framework, we envisioned creating a dedicated society for tractography, *not* merely another conference series, but a sustained platform where researchers could work together to solve the most pressing challenges in the field. This vision led to the formation of the International Society for Tractography (IST)**,** a nonprofit organization that unites neuroanatomists, developers, imaging scientists, clinicians, and end-users with a shared mission: to advance tractography as both a scientific discipline and a translational tool for human neuroscience and medicine. While centered on neuroanatomy, IST welcomes contributors from broader anatomical and biomedical fields relevant to tract-based imaging.

To launch this initiative, we organized the first Tractography and Neuroanatomy Retreat**,** informally known as the *Tract-Anat Retreat*, on the island of Corsica (https://tractography.io/tractanat_retreat/).

Bringing together 55 experts from Europe, North America, South America, Asia, and Australia (see Appendix Figure A1), the retreat was intentionally structured to break away from traditional conference formats. Instead of formal presentations or static posters, the retreat fostered dynamic, participatory engagement through roundtables, interdisciplinary discussions, and collaborative brainstorming sessions. The event was organized around three thematic tracks:

- Neuroanatomy and validation
- Tractography algorithms and modeling
- Clinical applications and translation

Each track was designed to highlight current strengths and limitations and identify collective challenges that could only be addressed through community-wide collaboration. The emphasis was on fostering critical discourse, encouraging participants to think beyond the scope of their own projects.

## Introducing the Millennium Pathways for Tractography

The most enduring outcome of the *Tract-Anat Retreat* was a shared desire to define a community-driven roadmap - a structured set of 40 grand challenges that should guide and shape tractography research over the next decade. Inspired by the *Millennium Prize Problems* in mathematics (Carlson et al. 2006), this effort crystallized into what we now call the **Millennium Pathways for Tractography,** as seen in Figure 1. The development of these challenges was rooted in the collective insights gathered during the retreat's thematic tracks. Through roundtable discussions and interdisciplinary collaborations, participants identified 40 grand short-, mid-, and long-term challenges spanning the full breadth of the tractography landscape. As seen in Figure 1, these challenges were then grouped into seven categories covering our three overarching themes of neuroanatomy, tractography, and applications:

A. Connectivity Characterization and Validation
B. Normative Modeling Across the Lifespan
C. Clinical Translation
D. Standardization, Accessibility, and Dissemination
E. Methods Development
F. Informatics
G. Linking Anatomy and Tractography

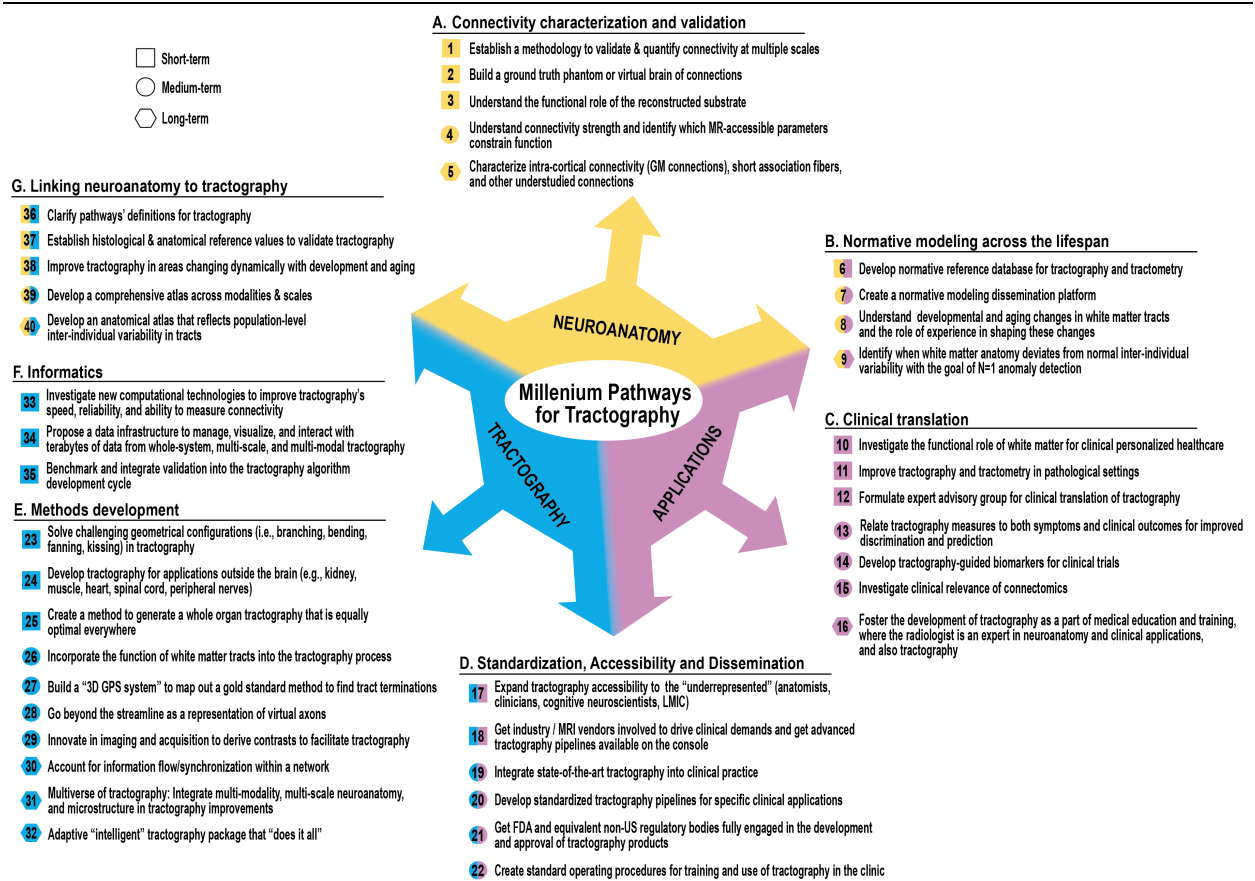

**Figure 1.** The Millennium Pathways for Tractography. 40 grand short-, mid-, and long-term challenges grouped into 7 categories across 3 overarching themes of neuroanatomy (yellow), tractography (blue), and applications (purple).

The Millennium Pathways for Tractography provides a framework to address fundamental questions and limitations in tractography while fostering methodological innovation, promoting rigorous validation, and advancing clinical applications. As a community-defined roadmap, these challenges aim to guide and inspire future work, ensuring that tractography continues to evolve in a manner that reflects both scientific rigor and translational relevance.

In the sections that follow, we present each of the seven Millennium Pathways for Tractography categories in detail, outlining the challenges they contain, the motivations behind them, and the opportunities they represent. Together, they offer a comprehensive view of where tractography stands—and where it needs to go.

# A. Connectivity Characterization and Validation (Neuroanatomy)

*What is connectivity, and how can we validate, recreate, and quantify it?*

Despite decades of progress in mapping the human connectome, the field still lacks a unified framework to define, quantify, and validate connectivity in the brain. These challenges span scales

- from single axons to large-scale networks - and aim to align imaging-derived connectivity with biology, function, and rigorous anatomical standards (Figure 2).

A fundamental *barrier is understanding how water diffusion measured by MRI relates to actual neuronal architecture*. Unlike microscopy, where manual proofreading can directly assess reconstruction fidelity, diffusion MRI provides only indirect cues. To move forward, we need gold-standard references - derived from histology, tracer studies, and high-resolution microscopy - that allow us to objectively "proofread" tractography, revealing where it succeeds, where it fails, and why. These benchmarks are essential for discovering generalizable rules that link diffusion paths to axons (*See Section G, Linking neuroanatomy to Tractography*), and for developing tractography algorithms that are not only technically valid but biologically grounded.

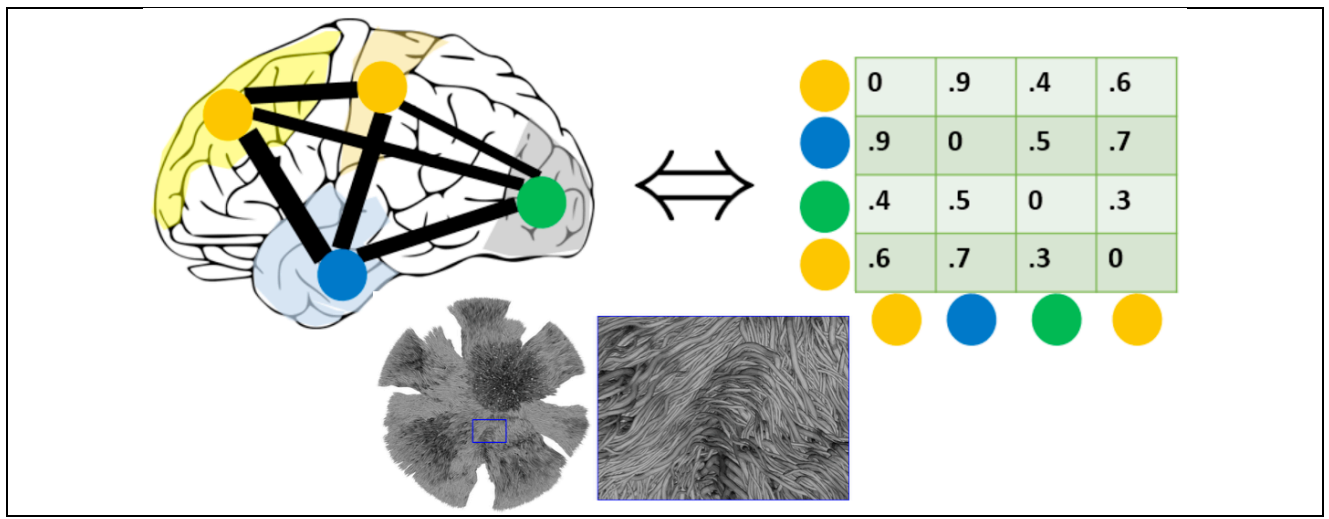

**Figure 2.** Five grand challenges of the *Neuroanatomy (yellow)* theme grouped into the *Connectivity Characterization and Validation* category.

## 1. Establish a methodology to validate and quantify connectivity at multiple scales [short-term]

*What is the methodology needed to validate tractography and quantitative connectivity?*

Tractography operates across spatial scales, yet our current tools often lack a scale-aware computational framework. Anatomically, at the smallest scale, connections involve individual axons, usually described by their origin, trajectory, terminations, diameter, and microstructure (Parent and Parent 2006). At intermediate scales, fiber bundles comprise collections of axons with shared geometry and/or targets (Zhong and Rockland 2003). At the macroscale, connectomes (Hagmann et al. 2008; Sporns et al. 2005) reduce brain connectivity to weighted graphs, often omitting biologically relevant detail.

A grand anatomical challenge is to define what constitutes a valid "connection", or a measure of "connection strength" at each of these scales (Dyrby et al. 2025), and to develop methods that allow us to measure and interpret them consistently. This challenge encompasses two related concepts: determining if a pathway exists is often treated as a binary question (a 'connection', linked to geometric accuracy), while quantifying its properties is a continuous or ordinal (Stephan et al. 2001) measure ('connection strength' or 'connectivity'). Regardless of whether we consider connections as binary, ordinal, or continuous, a fundamental task is to define these concepts and measure them consistently. Once defined, the tractography challenge is to find what metrics (e.g., streamline count, density, microstructural composition) meaningfully reflect connection strength. How do these metrics translate across scales and how different metrics interrelate, and under what assumptions.

There is also a need for multiscale validation tools (Kjer et al. 2025). Existing benchmarks (Thomas et al. 2014; Maier-Hein et al. 2017; Cote et al. 2013; Lawes et al. 2008) rarely extend beyond local geometric accuracy, and most validation studies are limited to N=1 or highly selected samples (due to cost, etc.). Advancing beyond this will require systematic comparisons across species, imaging modalities, spatial resolutions, and more evident parallels of what tractography should be expected to resolve at each level. This challenge is closely linked to *Challenge #37* at the intersection between tractography methods and neuroanatomy themes.

## 2. Build a ground truth phantom or virtual brain of connections [short-term]

*How do we build brains to validate tractography? What number(s) should be put in the connectivity matrix?*

Robust validation requires reference data (Hayashi et al. 2021), yet the field still lacks a widely accepted *ground truth* for structural connectivity, as also just described in *Challenge #1*. *Ex vivo* anatomical methods such as tracer studies or Klinger dissection (Martino et al. 2010) are labor-intensive and hard to scale (*Challenge #37*).

A key goal is to develop a "virtual brain" or multimodal phantom that combines high-resolution microscopy, anatomical dissection, and diffusion imaging data. Such a dataset would offer a common validation platform and allow comparison of tractography outputs against known anatomy. Ideally, a collection of these "virtual brains" should be created, incorporating data from multiple brains or tissue blocks to characterize inter-individual variability, span regions of differing complexity (cortex, brainstem, deep GM), numerous bundles, and at various scales. This effort will also require consensus on data formats, nomenclature, and what constitutes sufficient evidence for a tract to be considered "real" - echoing *Challenge #1* (defining a valid connection), and linking closely to Challenge #36 which emphasizes the need for standardized pathway definitions and nomenclature conventions.

### 3. Understand the functional role of the reconstructed substrate [short-term]

*How do we factor in the functional role of WM?*

Recently, Gordon et al (Gordon et al. 2023) refined the famous motor homunculus in a continuum of functions at the cortical level. However, it is still unclear whether distinct white matter pathways correspond to the functional regions observed in the cortex, and how white matter contributes to shaping or modifying these functions. Mapping functional specialization across cortical territories has been extensively investigated using invasive and non-invasive functional neuroimaging methodologies. However, the relationship between structural connections, brain function, and functional outcomes remains poorly defined. This complexity has been formalized in recent network neuroscience models, which describe how structural pathways shape, but do not determine, functional dynamics (Avena-Koenigsberger et al. 2017; Seguin et al. 2023). The assumption that "structure supports function" is widely accepted, but still lacks specificity - particularly in light of perspectives that emphasize that white matter (its conducation properties, and spatial relationships or communication with other cells) actively shapes neural computation and should be considered integral to brain function rather than peripheral to it (Bullock et al. 2005).

To advance our understanding of the white matter's role in functional processing, it is necessary to delineate which structural features - such as axon diameter, myelination, path geometry (length, cortical coverage, shape) or those potentially unrecoverable with dMRI, like local circuit integration, dendritic architecture, direction of signal propagation, or synapse density – most directly influence signal transmission. Do these features affect the timing, bandwidth, synchronization, or modulation of neural communication (Fries 2005; Innocenti et al. 2015; Fries 2015)? Further, can tract-level microstructural properties account for inter-individual variability in cognitive performance or behavioral phenotypes? These questions form the foundation for targeted investigations that bridge anatomical structure and functional outcome. This challenge is also addressed in Challenge #10 (*Investigate the functional role of white matter for clinical personalized healthcare*).

### 4. Understand connectivity strength and identify which MR-accessible parameters constrain function [mid-term]

*What biophysical properties underlie connection strength, and can they be measured with MRI?*

To support quantitative models of brain function, we must move beyond binary notions of "connected" versus "not connected" representations and begin quantifying connectivity. While this concept remains loosely defined, potentially encompassing axon count, myelin density, conduction velocity, or functional synchrony, its refinement is critical to bridging structure-function relationships and relates directly to Challenge #1, above.

Developing tract-level biomarkers that quantify connectivity will require both progress in anatomical clarity and technological tractography (and more broadly, MRI) innovation. Key challenges include identifying which features can be meaningfully estimated from MRI, validating them against independent modalities, and understanding their variation across individuals, populations, or disease states. Ultimately, the value of any proposed "connectivity" metric will be judged by its utility (e.g. prediction ability). A critical benchmark should be its ability to inform and constrain models that can predict functional measures of connectivity (Siddiqi et al. 2022). Without this link, structural measures risk remaining mere descriptions of anatomical "scaffolding" rather than true bridges to understand brain function.

## 5. Characterize intra-cortical connectivity, short association fibers, and other understudied connections [long-term]

*How can we track understudied connections?*

Much of the tractography field has focused on long-range white matter bundles, yet intra-cortical and short-range connectivity accounts for most brain wiring. These include short association fibers (e.g. U-fibers), superficial white matter, and intracortical pathways, structures that support critical local processing, development, and plasticity. Yet, they remain poorly characterized and difficult to reconstruct (Shastin et al. 2022; Guevara et al. 2020; Van Dyken et al. 2024). Two major acquisition-related limitations are spatial resolution and image distortions. Spatial resolution often remains coarse relative to the size of these fine-scale structures, and distortions further complicate accurate anatomical mapping and precise spatial localization. Anatomical challenges are equally important: these pathways differ markedly from deep white matter, exhibiting lower anisotropy due to lower axon density, reduced myelination, and increased fiber dispersion (both within and near the cortex). Most tractography methods (and reconstruction algorithms) were developed for deep white matter and may not translate well to these environments. Finally, the anatomy of these pathways remains poorly and inconsistently studied, resulting in a lack of precise definitions and consensus dissection protocols that impede both reproducibility and algorithmic training. Together, these limitations explain why such structures remain largely absent from the tractography literature. Imaging strategies specifically optimized for small, thin, short, or highly curved bundles will be crucial in the future to address this challenge (Schilling et al. 2025).

This "dark matter" of connectivity, i.e. connectivity within the gray matter (deep gray matter or cortex) is ubiquitous yet largely invisible. Gray matter (GM) volume changes a lot during the lifespan, as do MRI and diffusion MRI measures. It stands to reason that within GM connectivity would also change a lot. It may be key to understanding how microcircuits contribute to cognition, behavior, and vulnerability to disease. Like cosmological dark matter, these short-range pathways are challenging to detect and may hold crucial clues about the brain's complex connectivity and function (Dumoulin 2017; Assaf 2019). Advancing this frontier will require tailored imaging protocols, refined modeling strategies, and cross-modal anatomical validation.

## B. Normative modeling across the lifespan (Neuroanatomy & Applications)

*Can we create normative references to the evolution of WM and its multiple tracts across age?*

Tractography offers a non-invasive window into WM structure, yet its application across the human lifespan remains underutilized. To enable meaningful personalized applications, it would be highly valuable to establish normative references, clarify developmental and aging change trajectories, and define deviation from typical variation (Marquand et al. 2019; Rutherford et al. 2022). In parallel, achieving this goal requires that tractography itself be treated as a quantitative measurement tool - requiring careful consideration of which features are meaningful, interpretable, and robust enough to serve as phenotypes that can be modeled across individuals and time. This section outlines the grand challenges to realizing tractography as a robust tool for lifespan neuroscience and precision medicine (Figure 3).

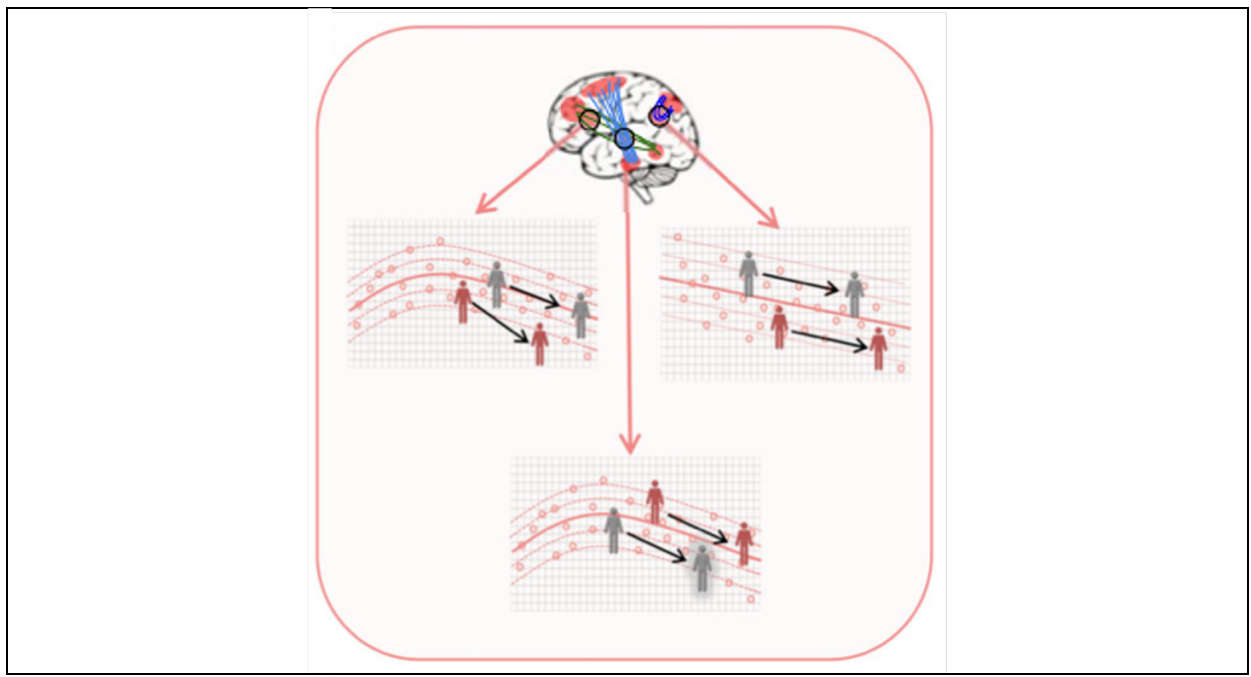

**Figure 3.** Four grand challenges at the intersection of the *Neuroanatomy (yellow)* and *Application (purple)* themes grouped into the *Normative modeling across the lifespan* category. The figure illustrates 3 white matter tracks of interest and 2 individuals (red and gray persons) that evolve differently in time.

## 6. Develop normative reference database for tractography and tractometry [short-term]

*Can we develop a normative reference of specific tracts and MR-accessible metrics across the lifespan?*

A foundational step toward tract-based precision neuroscience is the creation of large-scale, publicly accessible reference datasets that define how white matter properties evolve across the lifespan (Figure 3). Just as pediatricians rely on normative growth charts, we need standardized models for the trajectories of white matter bundles from infancy through old age. These should include tract-level microstructural (conventional DTI metrics as well as advanced metrics such as axon diameter, myelin water fraction, quantitative magnetization transfer, amongst others) and macrostructural (bundle shape and volume) metrics all measured across diverse populations and imaging conditions, as conceptualized and shown in Figure 3.

Such references must be built on harmonized acquisitions, robust tractography pipelines, and statistically sound modeling frameworks to ensure clinical and research utility. A parallel grand challenge, however, is to develop robust post-hoc harmonization techniques; such tools would be invaluable for integrating legacy and multi-site data, allowing them to be meaningfully compared against these normative references despite originating from non-harmonized acquisitions. Together, these frameworks should also account for sources of biological variability (e.g., sex, genetics, comorbidities) and technical variability (e.g., scanner, protocol). The goal is to establish a baseline against which individual measurements can be compared in both research and clinical contexts, enabling the identification of atypical development or degeneration.

## 7. Create a normative modeling dissemination platform [mid-term]

*How can we deliver personalized, tract-level insights from normative models to the broader community?*

Reference models are only as useful as their accessibility. We can draw inspiration from established open-access platforms designed to chart normative grey matter development and structure (Muili et al. 2024; Ge et al. 2024; Bethlehem et al. 2022). To operationalize their use, we must develop an open, flexible platform for visualizing, comparing, and integrating normative white matter data (see Figure 3). This includes intuitive interactive tools that allow users to upload individual data, receive tract-level deviations, and interpret results in the context of population-level distributions. To support interoperability with other types of data and broad use, application programming interfaces (APIs) to these models - the software building blocks of applications that use them - should enable integration into existing research pipelines and clinical platforms.

Such a platform would allow easier deployment to support clinical research, and may ultimately support future clinical applications clinical applications (e.g., detecting abnormal maturation in pediatric populations or accelerated decline in aging cohorts) and foster standardized reporting of white matter phenotypes across studies. Beyond single-metric comparisons, the platform should support the derivation of multivariate white matter phenotypes - composite signatures reflecting disease processes, developmental stages, or cognitive capacity. These "white matter fingerprints" could offer new biomarkers to detect anomalies (Chamberland et al. 2021) - for disorders such as Alzheimer's disease, schizophrenia, or traumatic brain injury, complementing other imaging modalities. Importantly, this infrastructure must also be adaptable as new data are acquired and modeling techniques evolve.

## 8. Understand developmental and aging changes in white matter tracts and the role of experience in shaping these changes [mid-term]

*Can we understand the role of experience in shaping developmental and aging WM changes?*

White matter development is not uniform. Different tracts mature at different rates (Lebel et al. 2008; Reynolds et al. 2019; Schilling et al. 2023; Yeatman et al. 2014; Brody et al. 1987) and are shaped by both intrinsic biology and external experience, which is also addressed in Challenge #25 (*Create a method to generate a whole organ tractography that is equally optimal everywhere*). Similarly, degeneration patterns in aging vary by bundle, cortical region, and individuals. Understanding these typical and atypical trajectories requires longitudinal and cross-sectional modeling of structural changes over time (see *Challenge #7* above).

Experience-dependent plasticity also plays a role. Learning, injury, stress, disease, and environment can alter white matter development and recovery (Knowles et al. 2022a; Knowles et al. 2022b; Huber et al. 2018; Sampaio-Baptista and Johansen-Berg 2017). Emerging research on the microbiome and gut-brain axis suggests these biological systems may also influence white matter plasticity (Loh et al. 2024). Tractography must evolve to capture not just static anatomy but also the dynamic, adaptive properties of white matter, such as pruning, myelination, or changes in tract geometry. This is also covered by challenge #38 below (*Improve tractography in regions that change dynamically (development, aging, etc.)*).

Moreover, white matter tracts do not evolve in isolation. Their development and degenerative trajectories are shaped by their interactions with other tracts, the gray matter regions they connect, and the broader functional networks in which they participate. Capturing these interdependencies will require integrated models linking microstructure changes to network-level changes. Moving forward, tractography must support not only localized measurements but also systems-level inferences: enabling us to ask not just *how* a tract changes, but *why* it does so, and *what consequences* these changes have for cognition and behavior.

## 9. Identify when white matter anatomy deviates from normal inter-individual variability with the goal of N=1 anomaly detection [long-term]

*How can we detect when an individual's white matter anatomy deviates from normative variation?*

The goal of normative modeling is to enable subject-specific assessments. This requires robust statistical models of tract variability that distinguish typical anatomical variation from clinically meaningful deviations. The key challenge is defining thresholds for anomaly detection – especially in the absence of ground truth – and ensuring that these are interpretable and actionable, as seen in Figure 3.

Subject-specific detection must also be context-sensitive. What constitutes a pathological deviation in one age group or clinical population may be normative in another. Ultimately, tractography should support individualized assessments (N=1 modeling) that inform diagnosis, prognosis, or intervention. Achieving this level of specificity requires both algorithmic innovation and conceptual clarity. Precision tractography is not about building custom pipelines for every subject, but about designing tools that can recognize when, where, and how an individual's anatomy deviates from normative expectations in a clinically meaningful way. While this vision aligns with broader themes across anatomy, methodology, and application, it also forms the foundation of what we describe in Challenge #32 (*Tractography that does it all, and adapts to the user, for a personalized algorithm and personalized results*).

## C. Clinical translation (Applications)

*How can we bring quantitative tractography into the clinic?*

While tractography has become a staple of academic neuroimaging, its clinical translation remains limited. At the same time, emerging clinical use cases – especially in neurosurgical planning, deep brain stimulation (DBS), traumatic brain injury (TBI), and demyelinating diseases – highlight its potential value (Bizzi et al. 2025). Addressing the challenges outlined in this section would move tractography beyond research pipelines and into routine clinical use. Doing so requires not only better algorithms but also infrastructure, education, validation, and engagement with regulatory and clinical communities (Figure 4).

### C. Clinical translation

- Short-term
- Medium-term
- Long-term

10. Investigate the functional role of white matter for clinical personalized healthcare
11. Improve tractography and tractometry in pathological settings
12. Formulate expert advisory group for clinical translation of tractography
13. Relate tractography measures to both symptoms and clinical outcomes for improved discrimination and prediction
14. Develop tractography-guided biomarkers for clinical trials
15. Investigate clinical relevance of connectomics
16. Foster the development of tractography as a part of medical education and training, where the radiologist is an expert in neuroanatomy and clinical applications, and also tractography

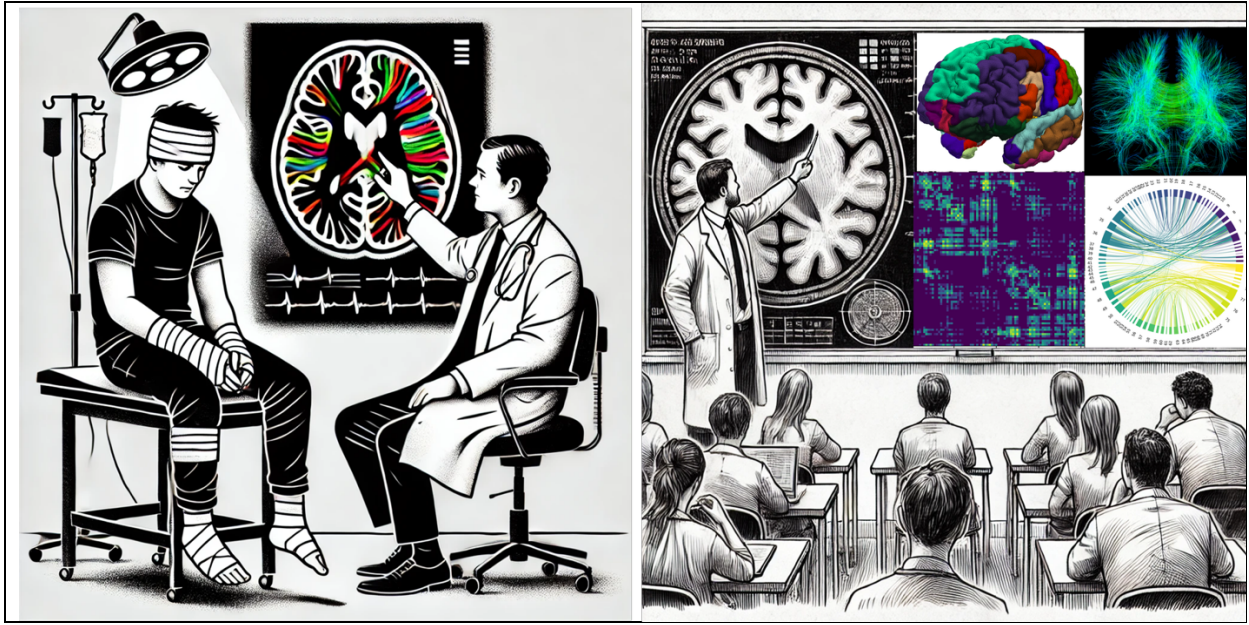

**Figure 4.** Seven grand challenges of the *Application (purple)* theme grouped into the *Clinical translation* category. Left panel illustrates personalized medicine using using tractography and right panel shows tractography part of medical education and training, which were generated with AI and adapted.

## 10. Investigate the functional role of white matter for clinical personalized healthcare [short-term]

*How can we translate the functional role of white matter into tractography-derived biomarkers for clinical decision-making?*

A recurring theme throughout the *Tract-Anat* retreat was the need to move beyond structural description toward functional interpretation. In clinical contexts, anatomical reconstruction alone is rarely sufficient to assess brain health. To enable personalized healthcare, tractography must begin incorporating features that reflect the functional capacity of white matter pathways: their ability to conduct, synchronize, or modulate neural activity. This ambition builds directly on Challenge #3, which calls for identifying which structural or microstructural features most directly influence signal transmission, timing, and coordination.

Anatomically similar pathways may vary substantially in their capacity to carry information. These differences help explain individual variation in symptoms, prognosis, and treatment response. Building tractography-derived biomarkers that are predictive of function rather than merely descriptive of structure will require bridging anatomical features with functional outcomes through mechanism/biological insight, joint modeling, and normative modeling, among others.

## 11. Improve tractography and tractometry in pathological settings [short-term]

*Can we make tractography robust to pathology?*

Clinical translation cannot rely on methods optimized for healthy brains. In practice, tractography must deal with pathology (tumors, edema, lesions, surgical cavities, and neurodegeneration), all of which distort white matter architecture and challenge standard algorithms.

Robust tractography in these settings requires new approaches to model or dynamically adapt to altered tissue environments. Tracts may be displaced by mass effect, infiltrated by tumor, or degraded by demyelination, and resolving these scenarios is essential for both surgical planning and monitoring disease progression (Yang et al. 2021; Kamagata et al. 2024; Bizzi et al. 2025; Theaud et al. 2025). This challenge also includes tractometry, the quantitative measurement of tract properties in general and in the context of pathology. Accurate, localized measures of white matter change can inform clinical decision-making and help differentiate among overlapping disease presentations. A key open question remains: how do we compare the location of a pathway and the microstructure of a pathway across individuals or timepoints when that anatomy is no longer typical? These requirements align with the broader methods development goals outlined in Challenge #25 (*whole organ tractography that is equally optimal everywhere*), Challenge #26 (*incorporate function into white matter tracts into the tractography process*), and Challenge #32 (*adaptive intelligent tractography that "does it all"*).

## 12. Formulate expert advisory group for clinical translation of tractography [short-term]

*How can we coordinate clinical, technical, and regulatory expertise to translate tractography into practice?*

To advance clinical adoption, the field must establish a clear roadmap incorporating diverse expert perspectives and defining tractography's clinical role. A proposed solution is to assemble a clinical translation working group or advisory board composed of domain experts: domain experts (clinicians, neuroscientists) and technological experts including tractography developers where appropriate (if tractography does not require further development, an expert technical expert might suffice)

This group would define use cases, develop best practices, identify "maximally impactful applications", and provide guidance on regulatory pathways. It would also coordinate with existing clinical societies and regulatory agencies, ensuring that tractography methods meet both technical and clinical standards.

Efforts will likely be disease-specific or application-specific. The needs for neurosurgical planning differ from those in multiple sclerosis, epilepsy, or traumatic brain injury. One-size-fits-all solutions are unlikely to succeed. Instead, applications must be defined around specific pathologies, clinical workflows, and outcome measures.

## 13. Relate tractography measures to both symptoms and clinical outcomes for improved diagnosis and prognosis [mid-term]

*How can tractography metrics be linked to clinical outcomes?*

Clinical relevance depends on bridging the gap between image-based features and patient experience. Many studies correlate tractography measures with cognitive or motor symptoms – and important and clinically meaningful endeavor – fewer have extended these findings to longer-term clinical outcomes such as treatment response, survival, or functional recovery.

To maximize clinical utility, the field must move beyond correlation alone and ask whether tractography-derived features can inform diagnostic classification, prognostic expectations, or treatment selection. These goals require distinct validation strategies, but all stem from the same foundation: translating image-based metrics into actionable clinical insight. Each of these use cases - whether diagnostic, prognostic, or predictive - offers a distinct path to clinical impact. For instance, the ability to differentiate subtypes of disease (e.g., dementia) could itself be a clinically valuable application, even if it does not directly forecast disease progression. The distinction between prognostic and predictive is critical as prognostic biomarkers estimate likely disease course independent of intervention, while predictive biomarkers indicate likely benefit from a specific treatment or therapy. Both are essential for personalized medicine. This requires high-quality clinical data, robust statistical models, and prospective validation across centers. For example, in stroke rehabilitation, features and integrity of the corticospinal tract (CST) may be linked to motor outcomes/recovery. Tract-level information may help stratify patients who might benefit from insensitive therapy *vs.* those unlikely to recover function, informing care pathways and trial design. Tractography and white matter pathway-specific information may matter and are critical to investigate.

Clinical metrics should encompass both symptoms and endpoints. Not all symptoms are reversible, but tracking white matter change may help predict disease trajectory, stratify patients for trials, or personalize interventions.

### 14. Develop tractography-guided biomarkers for clinical trials [mid-term]

*How can clinical trials and specific therapies benefit from tractography-based biomarkers?*

Tractography offers a unique opportunity to define targeted, anatomically specific biomarkers (Sullivan et al. 2015; Shukla-Dave et al. 2019). These could inform trial inclusion criteria, stratify patients by risk, or serve as endpoints for assessing treatment response, particularly in disorders with known white matter involvement, which are numerous and widespread. However, tractography-based biomarkers remain rare, and most have not been validated across cohorts or linked to clinical utility. To advance their adoption, the field must define tract-specific metrics that are reproducible, sensitive, and interpretable. This includes specifying what features or contrasts should be extracted, how to quantify deviation from normative ranges, and how to relate these measures to underlying biology or treatment mechanisms.

Applications could include monitoring axonal integrity in cases of acquired brain injury, such as traumatic brain injury, the effects of brain tumors, and the impact of surgical trauma, mapping seizure propagation pathways in epilepsy, or identifying candidates for deep brain stimulation. Clinical trials provide an ideal setting to evaluate the feasibility, specificity, and prognostic value of tract-based biomarkers.

## 15. Investigate clinical relevance of connectomics [mid-term]

*Can tractography-based connectomics be used in the clinic?*

Connectomics extends tractography from individual bundles to whole-brain network analysis (Bassett and Sporns 2017). While it has demonstrated value in research, its clinical role remains largely undefined, with growing interest in network-guided neurosurgical planning (Duffau 2021; Martinez Lozada et al. 2025). Bridging this gap requires the field to identify network-derived metrics that are not only robust and reproducible but also clinically actionable, that may influence diagnosis, prognosis, or intervention.

This challenge asks the field to define clinically meaningful network metrics, explore their relationship to treatment planning, and identify when and how connectomics adds value. For example, can the connectome topology predict recovery from stroke? Can network disconnection guide neurosurgical decisions? Can alterations in network-derived graph statistics explain cognitive variability?

To realize clinical utility, connectomics must first establish anatomical validity. At present, many connectivity matrices remain technically obscure - dependent on numerous modeling assumptions and sensitive to processing pipelines. Without clear anatomical validity, their clinical interpretability is limited. Progress will require robust validation of network edges (connections), demonstration of anatomical plausibility, and reproducibility across subjects and sites. Only then can we move beyond exploratory correlations to show that connectome-based measures add unique value in clinical settings. This includes designing studies that benchmark connectomic metrics against conventional predictors, and translating validated findings into tools that inform diagnosis, prognosis, or intervention.

## 16. Foster the development of tractography as a part of medical education and training [long-term]

*How can tractography, 3D, and connectional white matter anatomy be integrated into medical training?*

For tractography to be reliably adopted, clinicians must understand both its potential and its limitations. This requires incorporating tractography into the core education of radiologists, neurologists, and neurosurgeons - not as an optional specialty, but as a standard component of neuroanatomy and imaging training. Radiology offers a useful precedent: modalities like angiography or perfusion imaging are now integrated into medical training and clinical interpretation. Tractography must follow a similar path, completed with defined competencies, teaching resources, and certification pathways.

Such training should include (1) the physics and assumptions behind tractography, (2) common failure modes and artifacts, (3) interpretation of tract-based reports, and (4) clinical use cases and contraindications. Professional societies and academic programs can lead the way by offering tractography modules, case-based learning, and certification opportunities. This deep integration into current medical practice is key. Neurosurgeons (who are among the primary end-users of tractography and by for surgical decision-making) must be formally trained to critically evaluate and apply tractography results. Their mastery of neuroanatomy, imaging interpretation, and intraoperative planning makes them essential stakeholders in translating tractography into effective patient care. We can also look to radiologists, who routinely master complex imaging techniques. Their expertise in image arrtifacts and interpretation provides the foundation needed for tractography. By incorporating tractography into their training - much like their training in angiography or perfusion imaging - we ensure that these powerful tools are used with the necessary critical understanding, thereby enhancing, rather than replacing, existing professional roles and standards.

## D. Standardization, Accessibility and Dissemination (Tractography & Applications)

*How do we standardize, communicate, disseminate, and make tractography more accessible?*

A key take-home message from the Tract-Anat retreat was that we had many communication challenges, and we needed to talk more to each other.

The impact of tractography depends not only on the sophistication of its algorithms but also on the extent to which it is accessible, reproducible, standardized, clearly communicated (requiring consistent nomenclature), and trusted. Widespread clinical and research adoption demands a shift toward standardized pipelines, inclusive infrastructure, and transparent validation. This section outlines key challenges for ensuring that tractography is usable and valuable across a wide range of domains and settings, as seen in Figure 5.

**D. Standardization, Accessibility and Dissemination**

Legend:
- ☐ Short-term
- ◯ Medium-term
- ⬡ Long-term

17. Expand tractography accessibility to the "underrepresented" (anatomists, clinicians, cognitive neuroscientists, LMIC)
18. Get industry / MRI vendors involved to drive clinical demands and get advanced tractography pipelines available on the console
19. Integrate state-of-the-art tractography into clinical practice
20. Develop standardized tractography pipelines for specific clinical applications
21. Get FDA and equivalent non-US regulatory bodies fully engaged in the development and approval of tractography products
22. Create standard operating procedures for training and use of tractography in the clinic

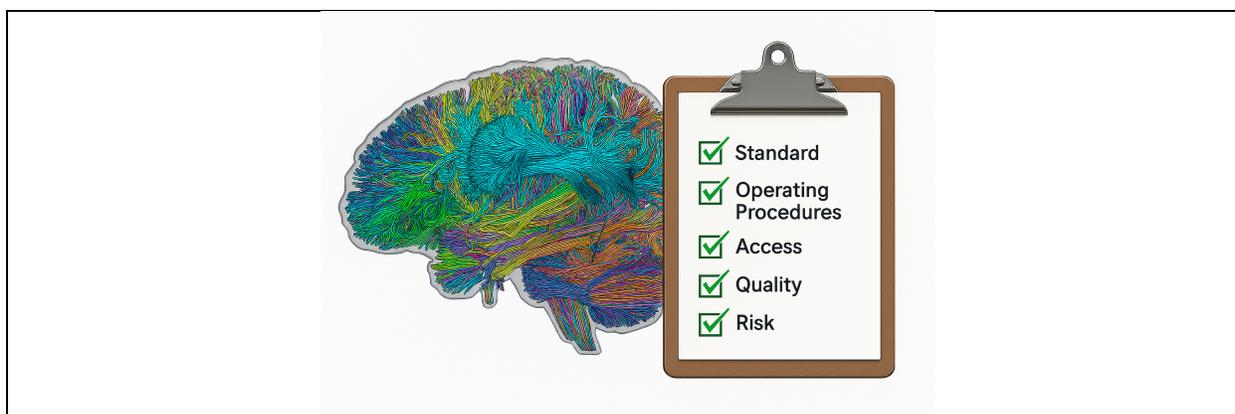

**Figure 5.** Six grand challenges of the *Tractography (blue)* and *Application (purple)* themes grouped into the *Standardization, Accessibility, and Dissemination* category. Illustrative concept of most challenges.

## 17. Expand tractography accessibility to the underrepresented (anatomists, clinicians, cognitive neuroscientists, low- and middle-income countries) [short-term]

*How do we increase the use of tractography to the whole world?*

Despite its relevance across disciplines, tractography remains underutilized by many potential user communities, including neuroanatomists, clinicians, and researchers in low- and middle-income countries (LMIC). For example, see the map of authors who submitted abstracts at the first IST conference, in Bordeaux 2025 (appendix Figure A2). Barriers include a lack of access to even basic imaging resources in some places, limited technical training, and insufficient communication between developers and applied users.

Bridging this gap will require simplified and robust tools, better outreach and interdisciplinary collaboration, streamlined software packages, curated educational content, high-quality and openly available datasets and targeted workshops that can help broaden participation. This also depends on infrastructure investments, including cloud-based processing and support for lower-field MRI systems.

## 18. Get industry / MRI vendors involved to drive clinical demands and make tractography pipelines available on the scanner console [short-term]

*How do we bring tractography into industry? How does industry tell us what clinicians and hospitals want?*

Tractography-based products are getting into the market, but for tractography to be adopted widely in clinical practice, it must move beyond standalone research tools and become integrated into commercial platforms. MRI vendors play a central role in enabling this shift. The field must work collaboratively with industry partners to define clinically relevant use cases, identify key technical requirements, and demonstrate tractography's value through outcomes data.

Deploying tractography simply and intuitively to clinicians is crucial for using tractography-derived features in real time. This transition will require simplification, automation, and the ability to convey not just pathway location but also reconstruction confidence, ideally through quantitative metrics or visualizations (Gillmann et al. 2021) that communicate uncertainty clearly.

## 19. Integrate state-of-the-art tractography into clinical practice [mid-term]

*How do we get advanced quantitative tractography methods into clinical practice?*

Even as research methods continue to advance, there remains a gap between high-performance tractography and clinically viable pipelines. Closing this gap requires more than technical optimization (optimizing tools for accuracy and usability, speed, and robustness across different acquisition conditions) it demands collaboration with clinical stakeholders to ensure tractography tools address real clinical needs and align with decision-making workflows. .

This includes listening to clinicians to understand when, where, and how tractography adds value to diagnosis, planning, or prognosis. Integration into clinical workflows requires pipelines that can be run with minimal manual intervention, interpretably segment key WM bundles, and return reproducible quantitative metrics. These methods must also be validated on clinical datasets, adapted to the practical constraints of routine imaging, such as shorter scan times and artifact-prone data.

Finally, practical deployment must consider the clinical infrastructure: results must be compatible with hospital systems and viewable within electronic medical records (EMRs), ensuring that tractography data is seamlessly incorporated into patient care alongside other diagnostic modalities.

## 20. Develop standardized tractography pipelines for specific clinical applications [mid-term]

*What is required to standardize tractography for specific clinical applications?*

No single pipeline will serve all clinical needs. Applications such as neurosurgical planning, epilepsy localization of invisible seizure focus, and deep brain stimulation (DBS) each place different demands on tractography methods. Building standardized, application-specific pipelines with harmonized acquisition, segmentation, and analysis protocols will support reproducibility and cross-site implementation.

Such pipelines should also include clear definitions of anatomical targets, quality control metrics, and reporting formats that align with clinical decision-making. In the long term, these standards will be necessary to evaluate tractography's role in clinical trials and therapeutic pathways. This effort depends on upstream consensus regarding relevant pathway definitions (see *Challenge #36*) and feeds directly into efforts to train and certify clinical users (*Challenge #22*).

## 21. Get FDA and equivalent non-US regulatory bodies fully engaged in the development and approval of tractography products [mid-term]

*What is required to gain regulatory approval for tractography-based clinical tools?*

While few tractography tools have historically been cleared for clinical use, recent regulatory approvals mark an encouraging shift. For example, advanced models such as constrained spherical deconvolution and probabilistic tractography have now been integrated into FDA-approved systems, including Brainlab Elements and Medtronic Stealth. These advances demonstrate how collaboration among vendors, clinicians, and researchers can successfully translate cutting-edge methods into clinical workflows familiar to neurosurgeons. Still, broader regulatory adoption of advanced tractography methods will required sustained engagement with agencies including the US FDA, European authorities (overseeing CE marking), and counterparts globablly. This involves putting quality management systems in place, clarifying intended use, establishing safety and efficacy criteria, and creating documentation that aligns with regulatory expectations. Ideally, this will be coordinated through collaborative efforts between academia, industry, and regulatory partners.

Establishing tractography as a regulated clinical product will require validation of both technical performance and clinical utility, which will depend on coordinated efforts between academia, industry partners, and regulatory experts. It also demands consensus or at least, best practices, across the field on tract definitions, performance metrics, and reporting standards. For example, in the USA, one viable path is to pursue 510(k) clearance through the use of predicate devices, demonstrating substantial equivalence to already-cleared imaging or post-processing software. However, this requires careful mapping between novel tractography outputs and recognized clinical endpoints or workflows. For more novel applications (e.g., predictive biomarkers), the De Novo pathway or complete PMA (premarket approval) process may be needed, necessitating even more substantial evidence of safety, efficacy, and clinical utility. Establishing tractography as a regulated clinical product will require not just algorithmic innovation, but regulatory strategy.

## 22. Create standard operating procedures (SOPs) for training and use of tractography in the clinic [mid-term]

*What are the SOPs for bringing tractography to the clinic? What are the guidelines and SOPs for teaching residents/medical students and giving them the tools to use quantitative tractography reports in clinical settings? Who creates these SOPs?*

Building upon the call in Challenge #16 to integrate tractography into core medical education, successful clinical deployment also hinges on standardized practices for both its application and training itself. This requires development of robust Standard Operating Procedures (SOPs). For clinical use, these SOPs should cover the entire workflow –from acquisition and processing, to tractography segmentation and generation of clinical reports - to ensure reliability and reduce variation across users and sites.

Education, as highlighted previously, is key to this challenge. Beyond simply including tractography in curricula (Challenge #16), SOPS are essential. This means defining structured teaching tools, creating comprehensive clinical case libraries, and establishing hands-on certification tracks. These standardized training programs will be vital to build confidence and ensure a baseline level of competence among all users, from medical students to practicing clinicians. Attendees of the Tract-Anat retreat believe the IST can play a key role in this educational challenge.

## E. Methods development (Tractography)

*How do we create better, faster, and improved methods?*

While tractography has become an essential tool for non-invasive structural connectivity mapping, the methods underpinning it remain limited in scope and accuracy. This section outlines challenges that span from modeling complex fiber geometries to developing adaptive algorithms that integrate anatomical knowledge and multimodal information. Each challenge reflects a critical bottleneck in the pipeline and an opportunity for transformative progress, as seen in Figure 6.

### E. Methods development

Short-term / Medium-term / Long-term

23. Solve challenging geometrical configurations (i.e., branching, bending, fanning, kissing) in tractography
24. Develop tractography for applications outside the brain (e.g., kidney, muscle, heart, spinal cord, peripheral nerves)
25. Create a method to generate a whole organ tractography that is equally optimal everywhere
26. Incorporate the function of white matter tracts into the tractography process
27. Build a "3D GPS system" to map out a gold standard method to find tract terminations
28. Go beyond the streamline as a representation of virtual axons
29. Innovate in imaging and acquisition to derive contrasts to facilitate tractography
30. Account for information flow/synchronization within a network
31. Multiverse of tractography: Integrate multi-modality, multi-scale neuroanatomy, and microstructure in tractography improvements
32. Adaptive "intelligent" tractography package that "does it all"

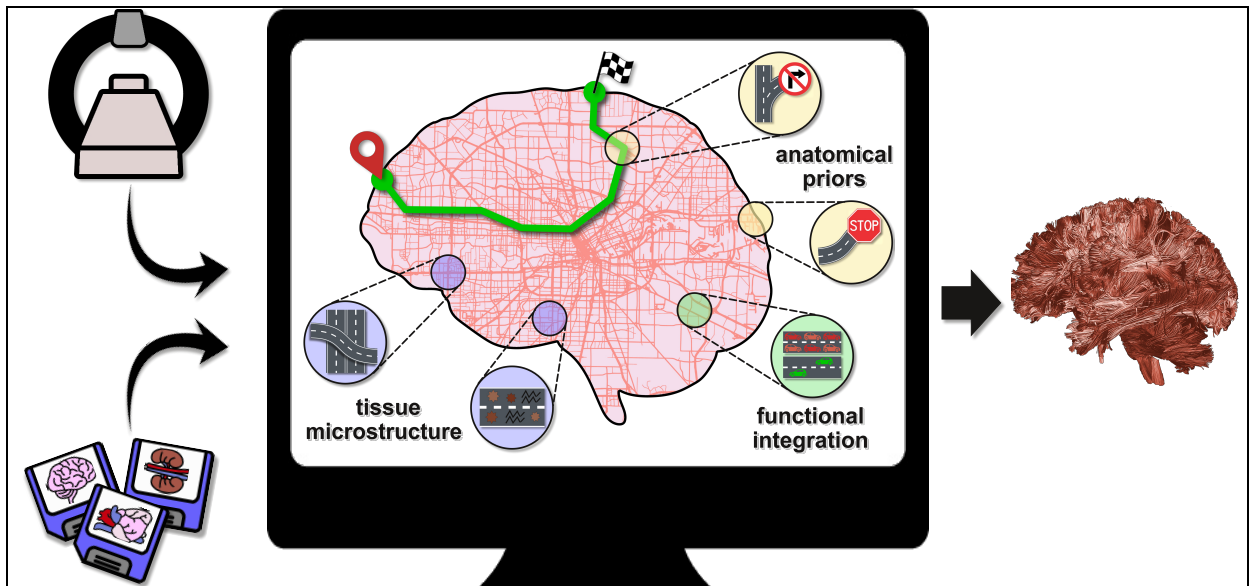

**Figure 6.** Eleven grand challenges of the *Tractography (blue)* theme grouped into the *Methods Development* category.

23. **Solve challenging geometrical configurations (i.e., branching, bending, fanning, kissing) in tractography [short-term]**

*How can tractography robustly reconstruct tracts irrespective of geometry complexity?*

Local tractography models still struggle to resolve fundamental geometric features of white matter: branching, bending, fanning, and touching fibers, which often result in errors or omissions. These configurations are prevalent throughout the brain and it is especially problematic for streamline-based tractography to correctly traverse these crossing fiber configurations. This challenge is as old as the diffusion tensor model (Basser et al. 1994; Pierpaoli et al. 1996) and the appearance of high angular resolution diffusion imaging techniques and models (Tuch et al. 2002). Disentangling them will require non-local approaches that have a more global context and more priors on the crossing fiber region to traverse (see Figure 6). Of course, a better understanding of the anatomical foundation (Kjer et al. 2025) on how axons crosses is crucial before solving the issue with methodological approaches.

24. **Develop tractography for applications outside the brain (e.g., kidney, muscle, heart, spinal cord, peripheral nerves) [short-term]**

*How can tractography be adapted and validated for use in non-cerebral tissues with distinct anatomical and biophysical properties?*

Tractography principles extend beyond the brain. Many systems - such as the spinal cord (a less studied part of the central nervous system), the peripheral nervous system (including cranial and peripheral nerves), and non-neural tissues like skeletal muscle, heart, and kidney - exhibit coherent microstructural organization that constrains water diffusion. These features can be leveraged to reconstruct structural pathways using tractography. Such applications offer powerful opportunities to visualize and quantify tissue architecture in both health and disease (Figure 6).

However, these tissues pose distinct challenges. Unlike cerebral white matter, anisotropy in peripheral tissues often arises from non-neuronal structures (e.g., myofibrils, tubules, fascicles) and is typically lower in magnitude. Additionally, organs like the heart or abdomen are subject to physiological motion, and tissue geometries may include sheets, laminae, or curved fascicles that violate standard streamline assumptions. To address the challenges, tractography outside the brain will require: Tailored acquisition protocols, including motion compensation, specialized coils, and optimized b-values; Modified modeling approaches, such as relaxed angular thresholds, low-anisotropy tracking, and multi-compartment models; and Organ-specific validation, leveraging histology, surgical findings, or gold-standard anatomical atlases.

Emerging applications include cardiac remodeling, spinal cord injury characterization, nerve pathway mapping, renal fibrosis assessment, and muscle architecture quantification. Advancing tractography beyond the brain offers not only novel research avenues but also clinically meaningful tools for tissue-specific diagnosis, surgical planning, and treatment monitoring.

## 25. Create a method to generate a whole organ tractography that is equally optimal everywhere [mid-term]

*How can tractography be equally optimal in all anatomical regions?*

Historically, tractography has focused on large, well-described, well-studied association, projection, and commissural white matter bundles. However, full-brain tractography must encompass challenging regions like the brainstem (see Figure 6), pons, cerebellum, and deep subcortical structures, areas with dense crossing fibers and high anatomical complexity. These regions are essential for sensorimotor function, consciousness, and autonomic control, yet they remain poorly reconstructed due to geometric, technical, and validation challenges. Tractography must also be equally optimal in the developing brain, where the reliability of fibers reconstructed depends on fiber density and myelination. The challenge is to be able to reconstruct fibers that are not yet mature, where orientations estimated from diffusion MRI are often more difficult to determine. Thus, tractography appears less efficient and reliable.

Current pipelines typically apply uniform parameters across the entire brain, which leads to suboptimal performance in anatomically diverse regions. Instead, tractography should adapt to local context, whether through bundle-specific thresholds, region-specific priors, or spatially varying model parameters. A long-term goal is to build a method that adapts its strategy based on anatomical location, producing whole-brain tractograms with consistently high fidelity across all regions.

## 26. Incorporate the function of white matter tracts into the tractography process [mid-term]

*How can functional priors or functional imaging be incorporated into tractography?*

Building directly on the need to quantify connectivity in a functionally meaningful way (Challenge #4), we must address how to incorporate the function of white matter tracts into the tractography process itself. While most current methods treat tractography as purely structural, with no regard for the functional role of tracts. However, the physiological function of a connection (conduction velocity, synchronization, or information flow) depends on microstructural features like axon diameter, myelination, and branching patterns. Incorporating functional relevance into the tractography pipeline will require joint structure-function acquisition beyond MRI, new modeling, and incorporation of prior anatomical knowledge into the problem.

## 27. Build a "3D GPS system" to map out a gold standard method to find tract terminations [mid-term]

*How can we inform tractography with all the multi-modality and a prior information to help guiding?*

Determining where streamlines should begin and end remains a major source of error. A "3D GPS system" would integrate coordinate frameworks, anatomical constraints, and probabilistic models to guide streamline terminations with higher anatomical accuracy, analogous to route planning in GPS systems, which account for known road networks, destinations, and travel rules. Tractography algorithms would need to integrate rules such as "turn left in 50mm" or "bottleneck ahead", based on probabilistic waypoints from tract templates, priors, and atlases. Such GPS operations are shown in Figure 6, illustrating the concept of GPS-rules that would come from multimodal constraints (cortical parcellation, laminar priors, histological maps), analogous to start/end addresses in our GPS apps. Moreover, the tractography algorithm could also adapt dynamically during tracking to pause, reroute, and reweigh local evidence based on a combination of local and global rules. This grand challenge is closely related to long-term Challenges #31 (*Multiverse of tractography: Integrate multi-modality, multiscale neuroanatomy, and microstructure in tractography improvements*) and #32 (*Adaptive "intelligent" tractography package that "does it all"*) below.

## 28. Go beyond the streamline as a representation of virtual axons [mid-term]

*Can virtual axons have a different representation than the streamline?*

Streamlines are a useful abstraction, but they do not capture key biological features (axon diameter, branching, or topographic continuity) and may be misleading to those not familiar with the principles and limitations of tractography. Future work should explore richer geometrical and topological representations to overcome these limitations: bundles that reflect tract coherence and dispersion, sheets that model laminar or fan-like projections (Tax et al. 2016; Tax et al. 2017), and graphs that embed connectivity into mathematically structured networks. These representations may align more closely with known anatomy and facilitate better integration with complementary modalities such as tract tracing, high-resolution histology, or layer-specific imaging. Importantly, they may also better match clinical needs, where location, trajectory, and the spatial relationship between connections are often more actionable than abstract streamlines.

### 29. Innovate in imaging and acquisition to derive contrasts to facilitate tractography [mid-term]

*Can we find advanced diffusion MRI acquisitions or other non-diffusion MRI contrasts to help tractography?*

Despite algorithmic advances, tractography still relies heavily on contrasts provided by the diffusion MRI signal. Innovations in diffusion contrasts (new acquisition techniques, ultra-high-gradient MRI, oscillating gradients) and modeling (modeling orientation, anisotropies, cell sizes and types), non-diffusion contrasts (contrast mechanisms sensitive to myelin, axons, astrocytes, oligodendrocytes, microglia) such as elastography (density, viscosity of connections), vascular imaging, functional contrasts (WM BOLD), may provide complementary signals for resolving axon orientation, density, and microstructure. These efforts may be especially useful in distinguishing intra- vs. extra-axonal contributions or resolving small-scale fibers in deep or superficial regions.

### 30. Account for information flow/synchronization within a network [long-term]

*Can the dynamics and flow of information along white matter tracks help reconstruct them or disentangle the challenging ambiguities?*

While Challenge #26 explored incorporating functional relevance as a prior, this challenge delves deeper, focusing on the dynamics of information flow and the methodological hurdles involved. Tractography typically assumes undirected and static pathways, yet true brain function depends critically on timing**,** synchronization**,** and directionality. The core methodological question is twofold: first, can these dynamic aspects be estimated and modelled by integrating tractography with functional modalities (e.g., fMRI, MEG, neurostimulation)? Second, can insights into these dynamics - not just *where* connections are, but *how* and *when* they are used – be leveraged to refine tractography itself, potentially overcoming challenging structural ambiguities? Addressing these challenges could reshape how we interpret connectivity, moving us from a static map to a dynamic circuit diagram.

### 31. Multiverse of tractography: Integrate multi-modality, multiscale neuroanatomy, and microstructure in tractography improvements [long-term]

*How can we achieve the multiverse of tractography in one computational framework?*

The tractography field has grown increasingly modular, with separate tools for modeling diffusion, defining anatomy, and interpreting function. A grand challenge is to integrate these domains into a unified tractography framework that pulls from multimodal sources (e.g., histology, microscopy, MRI), spans scales (axon to brain), and incorporates priors from both anatomy and function. This integrative "multiverse" would provide a more complete, biologically grounded foundation for structural connectivity mapping. Imagine a tractography framework that does not operate in isolation, but instead synthesizes all available priors across scales and modalities. This grand challenge is also covered and partially addressed by Challenges #11 (*Improve tractography and tractometry in pathological settings*), #23 (*Solve challenging geometrical configurations (i.e., branching, bending, fanning, kissing) in tractography*), #24 (*Develop tractography for applications outside the brain (e.g., kidney, muscle, heart, spinal cord, peripheral nerves)*), #25 (*Create a method to generate a whole organ tractography that is equally optimal everywhere*), #26 (*Incorporate the function of white matter tracts into the tractography process*), #27 (*Build a "3D GPS system" to map out a gold standard method to find tract terminations*), #32 (*Adaptive "intelligent" tractography package that "does it all"*), and #38 (*Improve tractography in regions that change dynamically (development, aging, etc.)*).

Anatomical constraints derived from histology and comparative neuroanatomy (e.g., non-human primates or rodent based tract-tracing studies) could inform tract geometry and connectivity priors. Microstructural features (axon density, orientation dispersion, myelination) could be estimated from diffusion models and spatially anchored using cellular-resolution microscopy. Functional information (e.g., fMRI, calcium imaging, metabolic maps, electrophysiological recordings) would modulate tract relevance, directionality, or timing. Even non-neuronal elements, such as vasculature or glia, could shape the local context in which tracking proceeds. In such a system, white matter pathways are not mere visualizations, but serve as the structural lattice (*the backbone*) upon which the multiscale architecture of the brain is computationally assembled.

## 32. Adaptive "intelligent" tractography package that "does it all" [long-term]

*How do we get to a quantitative tractography tool that does it all, for all?*

The long-term vision is an adaptive, intelligent tractography framework: a method that dynamically adjusts to input data, anatomical region, scientific questions, and clinical situations. This system would incorporate multimodal priors, learn from past reconstructions, adapt to pathology or developmental stage, and generate outputs optimized for both research and clinical use. More than just an algorithm, it would be a tractography engine capable of generalization, adaptation, and guided inference across contexts. Therefore, this adaptive intelligent tractography would adapt to participants' metadata (age, sex, education, medical record, and more), leveraging all available digital data.

# F. Informatics (Tractography)

*How do we leverage the latest technology (GPU, AI, and others) to improve tractography?*

As tractography continues to evolve into a data-intensive discipline, it faces a new class of challenges related to scale, complexity, and integration. The field must now grapple with how to manage, interpret, and validate vast multimodal datasets, often spanning terabytes and multiple spatial resolutions. At the same time, there is an opportunity to harness emerging computational tools and architectures to improve speed, reliability, and interpretability. This section outlines key informatics challenges for building the next-generation tractography tools and infrastructure, as seen in Figure 7.

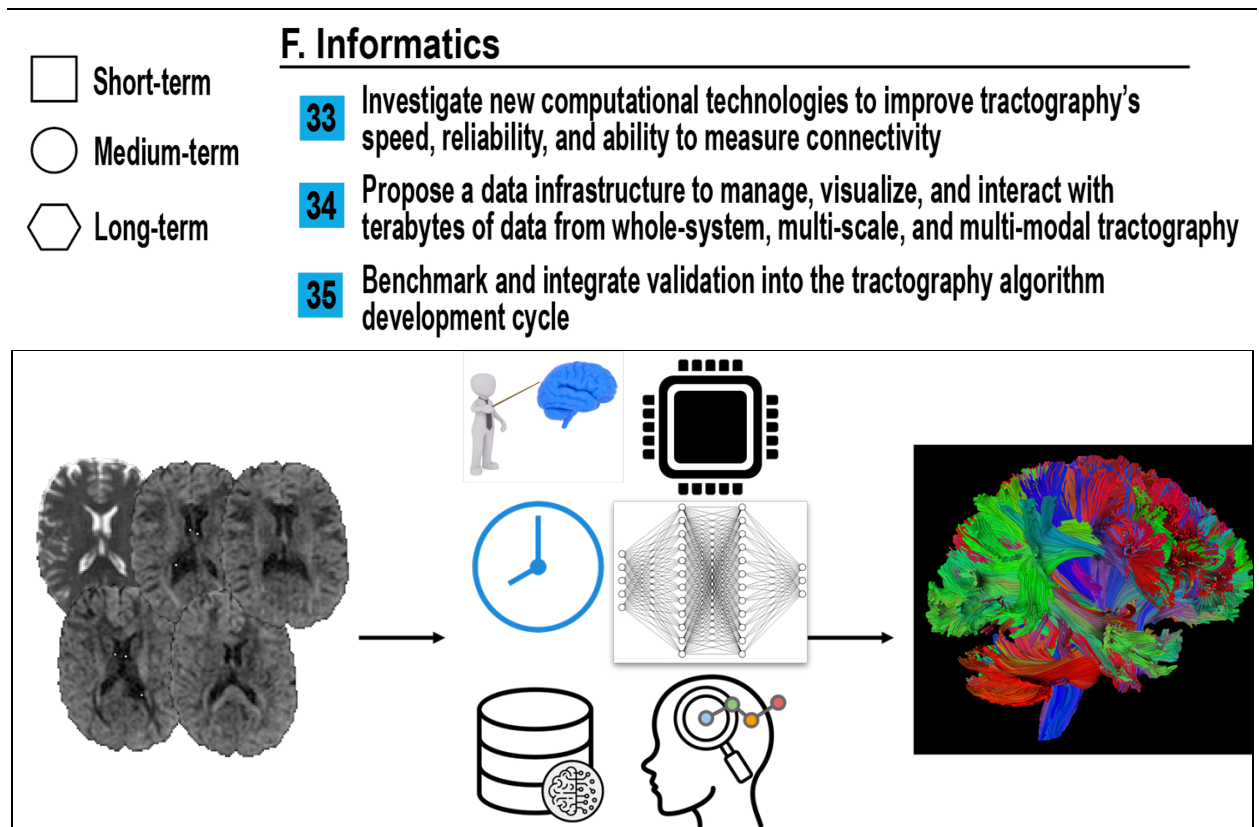

**Figure 7.** Three grand challenges of the *Tractography (blue)* theme grouped into the *Informatics* category.

### 33. Investigate new computational technologies to improve tractography's speed, reliability, and ability to measure connectivity [short-term]

*How can we leverage cutting-edge latest technologies to improve tractography?*

Modern advances in AI, GPU acceleration, cloud computing, and novel processing techniques have transformed other domains in imaging and data science. Still, tractography pipelines often remain slow, serial, and highly specialized. To scale up, tractography must integrate these emerging technologies, not just for faster streamline generation, but to improve anatomical accuracy and reproducibility, as illustrated in Figure 7.

Opportunities include:
- GPU-accelerated tractography pipelines for real-time or near-real-time applications.
- Machine learning frameworks for preprocessing (adaptive reconstruction, denoising, and artifact correction) and generative unsupervised and supervised tractography.
- Cloud-based pipelines that enable distributed processing and collaborative analysis.
- Exploratory research on quantum computing, leveraging its ability to perform complex calculations, to reduce the computing time of tractography algorithms.

These developments must be accompanied by rigorous benchmarking to ensure speed improvements do not compromise accuracy or anatomical validity.

### 34. Propose a data infrastructure to manage, visualize, and interact with terabytes of data from whole-system, multiscale, and multimodal tractography [short-term]

*How do we go from micron to meso to macro MRI scale and have references?*

Tractography has entered the petascale era. Even a single subject studied across these modalities may yield terabytes of information. As efforts scale to include large cohorts or multiple brain regions at cellular resolution, the demands on data infrastructure grow exponentially. Yet, we currently lack standard tools and frameworks to manage, explore, and share these complex, high-dimensional datasets across scales.

A priority is to develop an open and extensible **tractography informatics platform** (Hayashi et al. 2024; Markiewicz et al. 2021; Sunkin et al. 2013) that supports:
- Multiscale registration (e.g., light-sheet to MRI),
- Visualization of high-dimensional, multiresolution data,
- Efficient storage and data access protocols,
- Metadata standards for tissue source, imaging conditions, and annotations,
- Scalable user interfaces for both algorithm developers and clinicians.

The vision is a "Virtual Brain Connection Observatory" where researchers can interactively explore, query, and benchmark white matter data, bridging anatomy, imaging, and computation. Similar to the Allen Brain Atlas, but centered around the full characterization of white matter - from the origin, trajectory, and termination of connections, to rich microstructural features such as axon diameter, myelination, white matter neurons, and local cytoarchitecture - integrating data from tractography, light sheet microscopy, neural tracers, mesoscopic neurophotonics (e.g., OCT/PLL), and macro-scale diffusion MRI, as shown in Figure 7.

## 35. Benchmark and integrate validation into the tractography algorithm development cycle [short-term]

*Can we include a system that incorporates benchmarking into the life cycle of tractography development?*

While Challenge #1 addresses the fundamental difficulty of defining what constitutes a 'valid connection' or 'ground truth', this challenge focuses on the crucial process of systematically evaluating how well our algorithms perform against such standards. Even with a perfect definition (which we lack), we still need a robust framework for testing. Currently, despite decades of method development, tractography lacks standardized tools for this validation. Too often, algorithms are evaluated only after deployment, validated in one area/standard, or on repeatability only, or not validated at all. Instead, we propose that validation and benchmarking must be integrated as a core, continuous part of the tractography development cycle, embedded in each iteration of modeling and deployment. This shifts the focus from solely defining "truth" (Challenge #1) to establishing a rigorous process for assessing performance against it. This will require:
- A tractography-specific benchmarking suite, with community-agreed metrics and test datasets.
- Curated challenges across a range of anatomical regions, species, and pathologies.
- Guidelines for "qualifying" a new tractography algorithm for specific clinical or research applications.
- Integration of these validation routines into software development workflows (e.g., continuous integration, test suites).

Such a framework can help identify not just what works, but under what conditions it works, and what failure modes remain.

## G. Linking Neuroanatomy to Tractography (Tractography & Neuroanatomy)

*How can we relate tractography-based reconstructions to anatomical ground truth?*

While tractography has proven invaluable in studying bundle/connectome and dysconnectivity syndromes, the missing link with anatomy is often an open question. This section addresses the need to bring tractography closer to known biology and known anatomy, using data across species, imaging modalities, and scales, as illustrated in Figure 8.

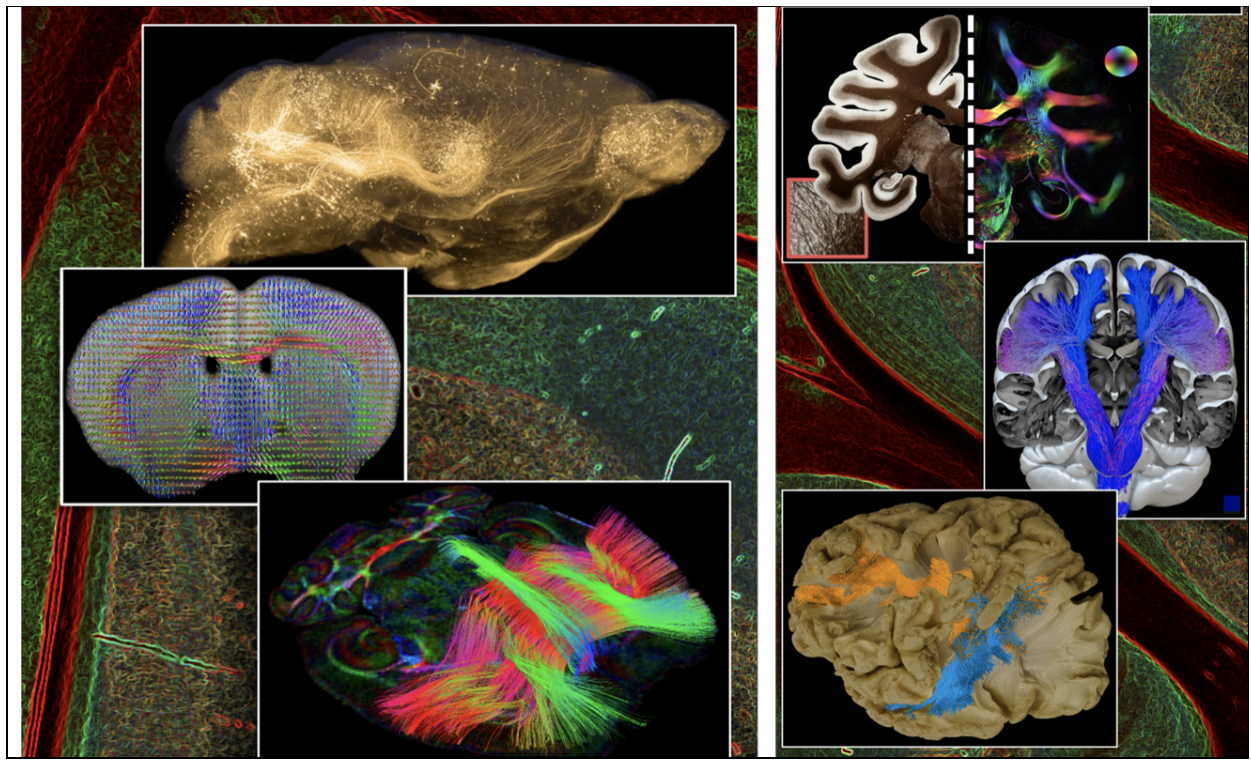

**Figure 8.** Five grand challenges of the *Tractography (blue)* and *Neuroanatomy (yellow)* themes grouped into the *Linking Neuroanatomy to Tractography* category.

## 36. Clarify pathways' definitions for tractography [short-term]

*How do we define white matter pathways that align with tractography's capabilities and constraints?*

A fundamental prerequisite for accurate tract reconstruction is a clear, operational definition of each pathway, encompassing its origin, termination, trajectory, and relative position. Yet this remains a major challenge. Definitions vary across historical sources and authors from historical cadaver dissections, species, and imaging modalities, and must now account for inter-individual variability and scale-dependent representations of anatomy.

Crucially, anatomical definitions must be resolution-aware: a useful tract definition must match the spatial and directional precision of the imaging modality on which it will be applied. Definitions that exceed the granularity of available data may be anatomically accurate in theory, but inapplicable in practice.

To move forward, the field must develop consensus definitions that are both biologically grounded and computationally tractable. This may require community-driven efforts to reconcile classical nomenclature with modern neuroimaging, define probabilistic or population-informed boundaries, and establish criteria for hierarchical segmentation. Emerging tools such as large language models (LLMs) may help digitize, synthesize, and standardize decades of anatomical literature.

Ultimately, these definitions will support reliable tract segmentation tools, reproducible atlases, and downstream applications in research and clinical contexts.

## 37. Establish histological & anatomical reference values to validate tractography [short-term]

*How do we establish reference values that can guide and validate tractography?*

Anatomical references are needed with sufficient resolution and biological accuracy to validate and guide tractography. Current resources are often incomplete, inconsistent across modalities, or limited to isolated regions or individual specimens. To be broadly useful, these reference datasets must sample multiple white matter tracts across the brain (not just well-studied structures like the corpus callosum) and characterize their diversity in geometry and microstructure. Importantly, histological features such as axon diameter, density, myelination, and orientation dispersion may vary by age, sex, or other biological factors. Since these datasets are necessarily derived from postmortem tissue, careful consideration must be given to selecting representative specimens and accounting for postmortem changes. Establishing tract-specific reference values, along with metadata on donor demographics and tissue processing, will be essential for comparing histology-based ground truth to in vivo diffusion-derived metrics (Figure 8).

This would allow for more rigorous benchmarking of novel methods and support model development. This effort must be paired with open-access repositories of co-registered histological and diffusion data, with expert annotations, which would also serve as invaluable tools for both validating algorithms and training machine learning models in the future. This closely aligns with Challenge #35 (*Integrating validation into the tractography development cycle*), Challenge #34 (*Propose a data infrastructure to manage, visualize, and interact with terabytes of data from whole-system, multiscale, and multimodal tractography*), and Challenge #2 (*Build a ground truth phantom or virtual brain of connections*).

## 38. Improve tractography in regions that change dynamically (development, aging, etc.) [short-term]

*How can tractography be robust in regions that change dynamically?*

White matter is not static - it undergoes dramatic remodeling during development, ages variably across tracts, and changes in injury and disease. Yet most tractography algorithms assume relatively fixed tissue properties, and most atlases reflect young adult or "average" anatomy. For example, some fibers are suppressed during early development (pruning). Hence, even if a tractography algorithm manages to reconstruct them, these reconstructed "to-be-pruned fibers" are often considered to be artefacts or noise because these fibers have no equivalent in the adult brain (Tovar-Moll et al. 2007; Benezit et al. 2015). Developing a tractography framework that can account for this across the lifespan is essential.

To better model dynamically changing brains, we need tractography pipelines that are sensitive to developmental trajectories, age-related microstructural shifts, and plasticity (Figure 8), leveraging the normative frameworks discussed in Challenge #6. For example, tracts exhibit remarkable changes in length, shape, volume, and microstructure (axon diameter, myelination) early in development, or may change trajectory as axons are pruned or reorganized. Similarly, aging introduces dispersion, atrophy, and lesion burden that alter fiber architecture and confound reconstruction.

A specific challenge here involves interpreting transient, yet potentially real, structures, such as those fibers pruned during development (Kostovic et al. 2019). Because they don't appear in adult references, they risk being misclassified as algorithmic errors. This highlights the need for new validation approaches, perhaps including a 'tractography reliability index'. Such a metric could quantify the algorithm's confidence in a reconstructed pathway, independent of standard adult anatomical priors, thereby helping clinicians distinguish genuine (though perhaps transient) features from noise – a critical step for applying tractography in dynamic conditions.

Robust solutions will require age-appropriate priors, tissue-specific modeling, and validation datasets spanning the lifespan. These efforts are tightly linked to Challenges #6-9, which focus on building normative references, platforms for detection, and personalized white matter phenotyping. These tools will be critical for pediatric imaging, aging research, and longitudinal disease progression or recovery studies.

## 39. Develop a comprehensive, multiscale brain atlas [mid-term]

*Can we create a brain atlas that integrates both macro- and microstructural information across scales and modalities?*

Despite the proliferation of white matter atlases (Wakana et al. 2004; Mori et al. 2009; Oishi et al. 2010; Guevara et al. 2012; Cho 2015; Yeh et al. 2018; Hansen et al. 2020), few provide true multiscale integration across modalities, species, and spatial resolutions. Most atlases focus either on gross anatomy (e.g., macro bundles) or high-resolution microstructural features (e.g., FA template), but rarely within a unified framework.

There is a critical need for comprehensive brain atlases that seamlessly integrate data from individual axons to whole-brain bundles, and align across multiple imaging modalities - including histology, light microscopy, and in vivo MRI. Ideally, such atlases would also incorporate information from comparative neuroanatomy to support cross-species inference and provide a common coordinate system for multimodal integration.

These atlases could act as a reference framework for both algorithm development and clinical interpretation, enabling meaningful comparisons across individuals, methods, and studies (Figure 8).

## 40. Account for variability across populations and time [long-term]

*Can we produce tractography atlases that capture meaningful variability across individuals, rather than rely on a single average brain?*

White matter anatomy is highly variable across individuals and the lifespan (due to development, aging, sex, handedness, pathology, and neurodivergence). Yet, most tractography methods and atlases are anchored to a single "average brain", often derived from a narrow sample of healthy young adults. This approach obscures biologically meaningful variation and limits sensitivity to individual differences.

One major goal is to quantify and model this variability, rather than treat it as noise. Doing so could improve the interpretation of individual scans, enable normative modeling, and enable more precise inferences about brain-behavior relationships. Achieving this will require integrating tractography with broader neuroimaging and behavioral data, as described in Challenges #6 through #10, #38, and #39.

Future atlases should move beyond static, single-population templates and instead adopt dynamic, probabilistic frameworks. One approach is the creation of continuous lifespan atlases, where age-dependent maps evolve smoothly from infancy to late life, enabling individualized comparisons at any time point. Another complementary strategy is implementing multi-atlas frameworks, drawing from populations that vary by age, sex, or clinical condition - an idea adapted from segmentation strategies in computer vision where multiple reference maps are fused to guide inference (Klein and Tourville 2012; Wang et al. 2013).

Together, these approaches would shift atlasing from a static tool to a flexible infrastructure for personalized, population-aware tractography.

# Conclusion

The Tract-Anat retreat brought together a diverse community to identify 40 "Millennium Pathways" for advancing tractography. These represent grand challenges across neuroanatomy, methods development, and clinical and scientific application, framed across short-, mid-, and long-term timelines. While not exhaustive, these pathways reflect a collective vision for where the field can and should go. They are intended as a living framework - open to revision, expansion, and refinement as the science evolves. This is a passport through the 21$^{st}$ century.

To realize this vision, sustained collaboration is essential. No single lab or domain can tackle these challenges in isolation. Progress requires integrating anatomical precision, advancing methodological innovation, and driving clinical and scientific relevance. This means anchoring tractography in biological ground truth, building validated and reproducible tools, and ensuring those tools address real-world questions in neuroscience and medicine. The International Society for Tractography (IST) is uniquely positioned to facilitate this progress by supporting working groups, hosting open discussions, and helping build the infrastructure needed to translate these ideas into practice.

These challenges represent a call to action: to deepen our anatomical understanding, strengthen our methodological foundations, and ensure that tractography delivers its promise for neuroscience and medicine in the decades ahead.

# Acknowledgements

Research Council of Finland, Grants #348631, #353798. Technology Industries of Finland Centennial Foundation, Grant #4302; Canada Research Chairs; MC acknowledges funding from the Dutch Research Council (NWO) grant numbers OCENW.M.22.352 and KICH1.ST03.21.004.; AD is supported by the Italian Ministry of Education, University and Research within the PRIN 2022 research program framework (#2022PXR8ZX); USherbrooke Research Chair in Neuroinformatics.; JD acknowledges funding support from the French government as part of the France 2030 programme (grant ANR-23-IAIIU-0010, IHU Robert-Debré du Cerveau de l'Enfant), the French National Agency for Research (grant ANR-22-CE37-0028), the IdEx Université de Paris (ANR-18-IDEX-0001), the Fondation de France and Fondation Médisite (grants FdF-18-00092867 and FdF-20-00111908).; This work was supported by the Donders Mohrmann Fellowship on 'Neurovariability' No. 2401515 (SJF) and the Dutch Research Council NWO Aspasia Grant 'Human individuality: phenotypes, cognition, and brain disorders' (SJF).; This research was supported by the Deutsche Forschungsgemeinschaft (DFG, German Research Foundation) - project number 222641018 - SFB/TRR 135 TP C10; CIHR PJT 470155; ANR-22-CE45-0004; AR's work was funded by National Institutes of Health grants MH121868, MH121867, R25MH112480, R01AG060942, and U19AG066567, and R01EB027585,as well as by National Science Foundation grants 1934292 and 2334483, and by the Chan Zuckerberg Initiative's Essential Open Source Software for Science program.; NIH R01NS125307; NIH K01EB032898 (K.G.S); SNS is supported by a European Research Council Consolidator Grant (ERC 101000969, NeuroMetrology) and the US National Institutes of Health Centre for Mesoscale Connectomics (NIH UM1NS132207 under the Brain Initiative).; National Institute on Aging grants U19 AG074866; Dr JYMY acknowledge position funding support from the Royal

Children's Hospital Foundation (RCHF 2022−1402), and support from The Kids' Cancer Project (TKCP) Col Reynolds Fellowship.; National Key R&D Program of China (No. 2023YFE0118600), the National Natural Science Foundation of China (No. 62371107); ANID-Basal AFB240002; NIH R01 NS095985; The research of ADL is supported by a Starting Grant from the European Research Council (agreement 101163214), the Galen and Hilary Weston foundation, and Stichting Hanarth Fonds.; NIH UH3NS103550, R01MH102238, R01MH132789, R01MH123542, Hope for Depression Foundation; The project has received funding from the European Research Council (ERC) under the European Union's Horizon Europe research and innovation programme (grant agreement No. 101044180, CoM-BraiN) (Principal Investigator: T.B.D.).; JDT by core funding from the Wellcome/EPSRC Centre for Medical Engineering [WT203148/Z/16/Z] and by the National Institute for Health and Care Research (NIHR) Clinical Research Facility at Guy's and St Thomas' NHS Foundation Trust. The views expressed are those of the author(s) and not necessarily those of the NHS, the NIHR or the Department of Health and Social Care.; Salary support from the NIBIB Intramural Program ; NIH R01MH125860, R01MH132610, R01MH119222 ; NSERC Discovery Grant RGPIN-2020-06109; M.T.d.S is supported by HORIZON- INFRA-2022 SERV (Grant No. 101147319) "EBRAINS 2.0: A Research Infrastructure to Advance Neuroscience and Brain Health", by the European Union's Horizon 2020 research and innovation programme under the European Research Council (ERC) Consolidator grant agreement No. 818521 (DISCONNECTOME), the University of Bordeaux's IdEx 'Investments for the Future' programme RRI 'IMPACT', and the IHU 'Precision & Global Vascular Brain Health Institute–VBHI' funded by the France 2030 initiative (ANR-23-IAHU-0001); Author C.M. has received support from the Brain & Behavior Research Foundation (BBRF) through the NARSAD Young Investigator Grant. ; This research was funded in whole, or in part, by a Wellcome Trust Strategic Award (104943/Z/14/Z), and Wellcome Discovery Awards (227882/Z/23/Z and 317797/Z/24/Z). For the purpose of open access, the author has applied a CC BY public copyright licence to any Author Accepted Manuscript version arising from this submission.

# Appendix

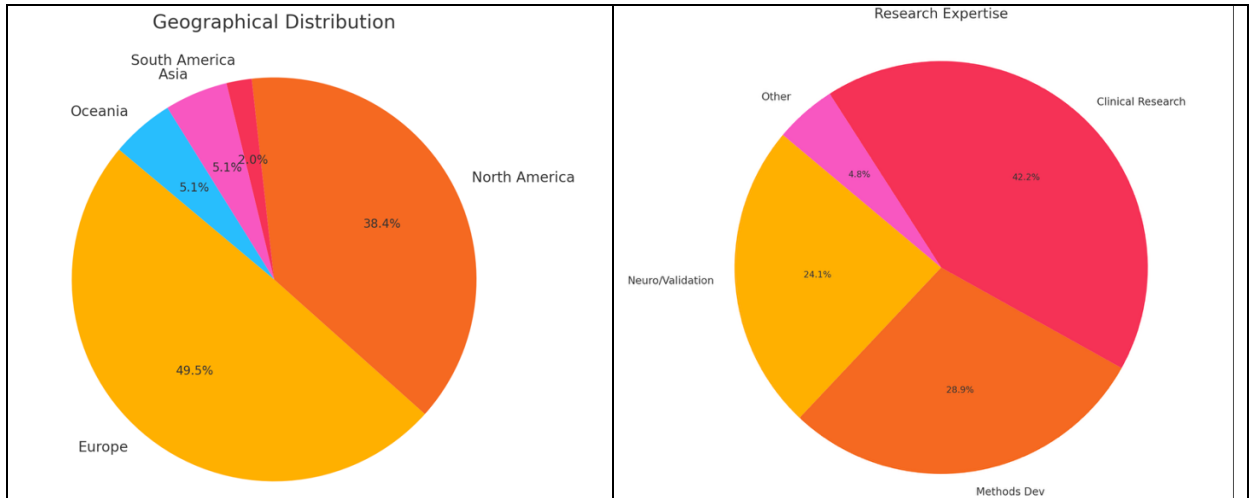

**Figure A1.** The Tract-Anat retreat initiative. Geographical and expertise distribution of participants. 33% of attendees were women. 49% from Europe, 38% from North America, 2% from South America, 5% from Asia, and 5% from Oceania. 42% of attendees identified themselves as clinical research scientists, 29% as tractography methods developers, 24% as neuroanatomy and validation experts, and 5% as other expertise, such as neurophonotics, network neuroscientists, and methods developers using tractography as input rather than methods for performing it.

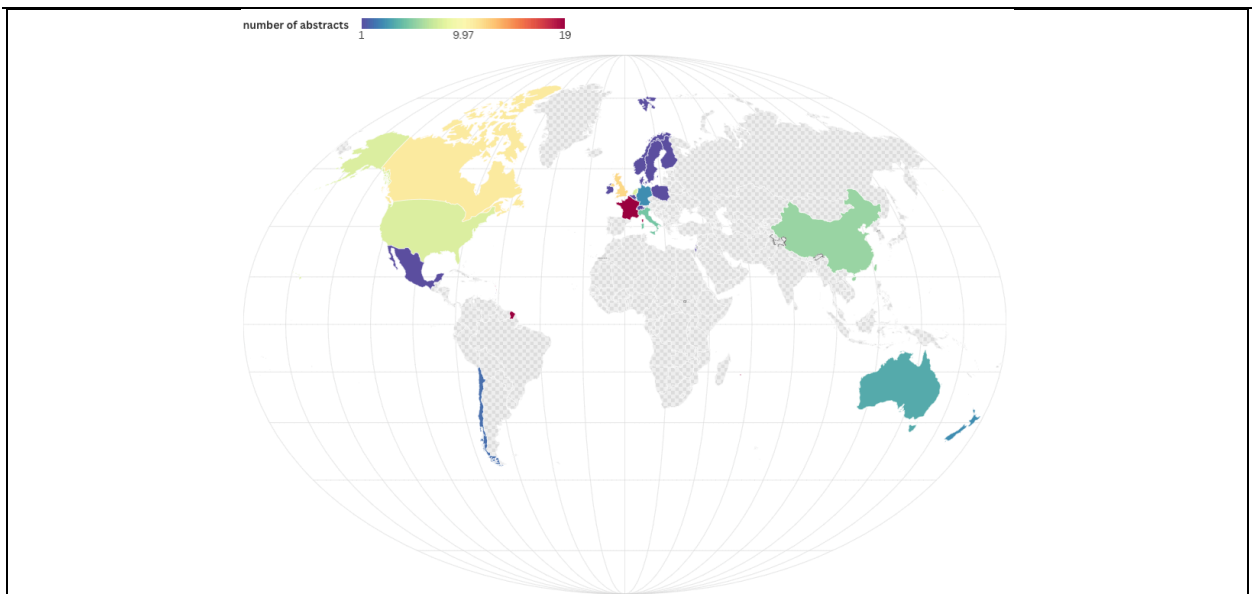

**Figure A2.** A map illustrating the geographical distribution of the 91 abstracts, from 21 different countries, submitted at the first IST conference, Bordeaux 2025.

## Structured Outputs and the *Brain Structure and Function* (BSAF) Collection

The *Tract-Anat Retreat* was more than an opportunity for scientific exchange - it was a structured experiment in collaborative output. Discussions were carefully scaffolded around key questions and shared goals, aiming to generate actionable insights and tangible contributions for the broader community.

Two structured activities formed the core of these outputs. The first was a series of "Did You Know?" sessions, where participants delivered concise, rapid-fire presentations highlighting surprising insights, misunderstood concepts, or unresolved technical and anatomical challenges. These sessions sparked lively debates and have since been distilled into a series of short papers aimed at correcting misconceptions and disseminating critical knowledge. Collectively, we discovered the great educational power of a single "Did you know" slide or short communication.

The second activity involved debate-style sessions that explored contrasting perspectives on unresolved methodological and interpretive issues, such as strategies for resolving crossing fibers, the merits of different validation techniques, and the appropriateness of various clinical applications. These debates, which paired advocates for opposing viewpoints, stimulated nuanced discussions that have also been synthesized into a forthcoming series of debate papers.

Significantly, these outputs are being consolidated and published within the Brain Structure and Function (BSAF) Collection [ref], providing the community with a permanent repository of curated insights. The BSAF Collection serves not only as an archive of the retreat's outcomes but as a catalyst for ongoing discussion and refinement of best practices in tractography.